\documentstyle[11pt,newpasp,twoside,epsf]{article}
\def\edcomment#1{\iffalse\marginpar{\raggedright\sl#1\/}\else\relax\fi}
\reversemarginpar

\begin{document}
\title{Fragmentation of a Protostellar Core: The Case of CB\,230}
\author{Ralf Launhardt}
\affil{Caltech, Astronomy Dept., MS 105-24, Pasadena, CA 91125, USA}

\begin{abstract}
The Bok globule CB\,230 (L\,1177) contains an active, low-mass star-forming 
core which is associated with a double NIR reflection 
nebula, a collimated bipolar molecular outflow, and strong mm continuum 
emission. The morphology of the NIR nebula suggests the presence 
of a deeply embedded, wide binary protostellar system. 
High-angular resolution observations now reveal the presence of two sub-cores, 
two distinct outflow centers, and an embedded accretion disk associated with 
the western bipolar NIR nebula. In terms of separation and specific angular 
momentum, the CB\,230 double protostar system probably results from core 
fragmentation and can be placed at the upper end of the pre-main sequence 
binary distribution.
\end{abstract}

\section{Motivation and observations}           

Binary systems have been observed in all pre-main-sequence 
stages of evolution and there is growing evidence for 
proto-binary systems, although the numbers are still small 
(Mundy, this volume).
Both theory and observations support the hypothesis that most binary 
systems form by fragmentation during the gravitational collapse of 
molecular cloud cores. 
We have started a program of high angular resolution observations 
to study the early formation process of binary stars in detail, 
using the Owens Valley Radio Observatory (OVRO) millimeter array. 

This paper presents new observations of the protostellar core in 
the Bok globule CB\,230 ($D$\,=\,450\,pc). 
CB\,230 contains a dense core (Launhardt et al. 1997, 1998, 2000) 
and two associated NIR reflection nebulae separated by $\sim$10\arcsec\ 
(Yun 1996; Launhardt 1996, Fig.1). 
The western nebula is bipolar, with a bright northern lobe 
perfectly aligned with the blue lobe of 
a collimated CO outflow (cf. Yun \& Clemens 1994, Fig.3). 
The eastern nebula is much fainter and redder 
and displays no bipolar structure. 
No associated stars are visible.

CB\,230 was observed at 1.2 and 3\,mm 
with the OVRO mm array in spring 2000. 
The mm continuum emission traced the optically 
thin thermal dust emission, while molecular gas was traced by the 
N$_2$H$^+$(1--0) and $^{13}$CO(1--0) lines at 93 and at 110\,GHz, respectively. 
N$_2$H$^+$(1--0) comprises seven 
hyperfine components and, compared to other molecules, 
depletes later and more slowly onto grains (Bergin \& Langer 1997). 
It is a reliable indicator of the morphology and kinematics of protostellar cores.
$^{13}$CO traces the small-scale structure of the outflow near the driving sources. 
Beam sizes for the different observations vary from 1\arcsec\ to 10\arcsec, 
and the spectral resolution was $\sim$0.2\,km/s.

\section{Results}                           

\subsection{Morphology and kinematics of the double core}   

\begin{figure} 
\plotone{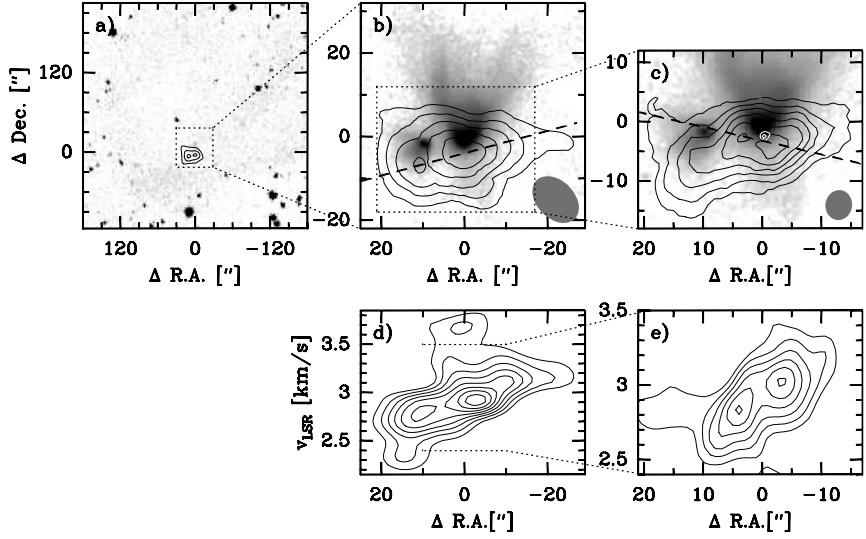}
\caption{\label{fig1}
\footnotesize
OVRO N$_2$H$^+$(1--0) observations of the CB\,230 double core. 
{\bf a)} Optical image (DSS) overlayed with N$_2$H$^+$\ contours. 
{\bf b)} K-band image of the NIR nebulae 
 overlayed with the intensity map of the 
 N$_2$H$^+$(1\,2\,--\,0\,1) triplet.
{\bf c)} As b) but at 4.5\arcsec\ resolution. 
 White contours show the 1mm continuum emission (cf. Fig.2).
 All contours start with 3$\sigma$.  
{\bf d)} and {\bf e)} Position-velocity diagrams of the 
 N$_2$H$^+$(1\,2\,3\,--\,0\,1\,2) transition along 
 PA indicated by the dashed lines in Figs. b) and c).
}
\end{figure}

The NIR morphology of CB\,230 suggests the presence of two 
deeply embedded YSO's separated by $\sim$10\arcsec\ (Fig. 1b). 
These were not resolved in previous single-dish mm observations. 
With higher angular resolution observations in a reasonably optically 
thin line which is unlikely to deplete in protostellar cores 
(N$_2$H$^+$(1-0)), the puzzle is solved.

At 9\arcsec\ resolution, the N$_2$H$^+$\ maps show that the molecular cloud core 
is elongated E-W and consists of two sub-cores separated in space 
(10\arcsec\ $\simeq$ 4500\,AU) and velocity. 
Each sub-core coincides spatially with one of the two NIR nebulae (Fig. 1b). 
The velocity field and the position-velocity (p-v) diagram along the 
major axis indicate that the double core rotates with 
$\omega$\,$\sim$\,2$\times$10$^{-13}$\,s$^{-1}$\ around an axis perpendicular 
to connecting line of the sub-cores and approximately parallel to the 
large-scale outflow axis (Figs. 1b and d). 
At higher angular resolution (4.5\arcsec), the stronger western sub-core splits up 
into two separate maxima located almost symmetric to the mm continuum 
source at the origin of the western outflow and NIR nebula (Fig.1c). 
Since there is also a large velocity gradient along this 
double-peaked western sub-core, it can be interpreted as a thick gaseous disk or 
torus with $R \sim 1600$\,AU, rotating at a rate of 
$\omega$\,$\sim$\,5$\times$10$^{-13}$\,s$^{-1}$\ around the embedded accretion disk 
(see Sect.2.2). 
Despite its smaller size, this torus has about the same specific angular 
momentum as the entire cloud core, 
$J/M$\,$\sim$\,5$\times$10$^{-17}$\,pc$^2$\,s$^{-1}$,  
suggesting that it has rotationally decoupled from the cloud core.

\subsection{Dust continuum emission from embedded disk(s)}     

\begin{figure} 
\plotone{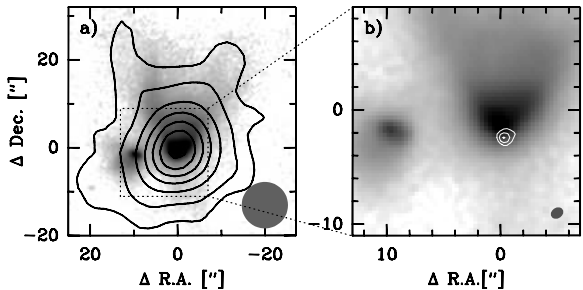}
\caption{\label{fig2}
 \footnotesize
1.3\,mm dust continuum emission from CB\,230. 
a) K-band image of the NIR reflection nebula (grey-scale) 
 overlayed with contours of the 1.3\,mm continuum emission 
 (IRAM 30-m; 1$\sigma$=7\,mJy/beam; levels 20, 40, 75, 105, 
 ... mJy/beam). 
b) Enlarged K-band image overlayed with the 1\,mm continuum emission 
 from OVRO (1$\sigma$=3.6\,mJy/beam; levels 11, 22, 33 mJy/beam). 
} 
\end{figure}

Strong 1\,mm continuum emission peaks at the 
origin of the western bipolar NIR nebula (Fig. 2a).
This dust emission is extended towards the eastern NIR nebula. 
The total mass of the CB\,230 core implied is 7\,M$_{\odot}$, of which 
$\sim$2\,M$_{\odot}$\ can be attributed to the unresolved
central condensation (Launhardt et al. 2000). 
Figure 2a explains the north-south asymmetry of the western bipolar nebula; 
the southern (redshifted) lobe is heavily obscured by dust in the dense 
core, while in the nortern (blueshifted) lobe we are looking directly 
into the outflow cone.
The OVRO 1\,mm dust continuum map shows a compact source associated with the 
origin of the western bipolar nebula and outflow (Fig.2b). 
It is unresolved in the 1\arcsec\ beam, 
suggesting the presence of an embedded accretion disk 
with radius $R$\,$<$\,200\,AU and mass $M$\,$\sim$\,0.1\,$M_{\odot}$.
A significant contribution by free-free emission can be ruled out since 
the bolometric luminosity of the entire cloud core of 11\,L$_{\odot}$\ 
points to a low-mass protostar with no capability to ionize its environment 
(Launhardt et al. 1997). The eastern source 
may be too faint to detect ($<$2\,mJy at 3mm and $<$10\,mJy at 1mm; 
$M$\,$<$\,0.006\,$M_{\odot}$) or simply lack a disk.

\subsection{The outflow is also double}    

\begin{figure} 
\plotone{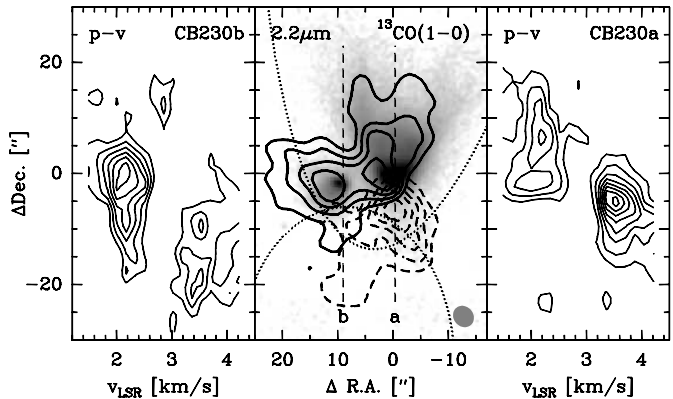}
\caption{\label{fig3}
 \footnotesize 
 OVRO $^{13}$CO(1--0) observations of the CB\,230 outflow.
 Middle panel: K-band image of the NIR nebula 
 overlayed with channel maps of the blue and red 
 $^{13}$CO(1--0) outflow lobes. 
 Blue lobe: $v_{\rm LSR} = 1.6\ldots 2.8$\,km/s, solid contours. 
 Red lobe: $v_{\rm LSR} =3 .0\ldots 4.2$\,km/s, dashed contours. 
 Contours start at 3$\sigma$.
 The dotted ellipses outline the large-scale $^{12}$CO outflow mapped by 
 Yun \& Clemens (1994; HPBW 48\arcsec). 
 Right and left panels: P-V diagrams for sources CB\,230\,a and b 
 along the slices indicated by dashed lines in the middle panel. 
} 
\end{figure}

The OVRO $^{13}$CO(1--0) map of CB\,230 at 4\arcsec\ resolution reveals 
that the large-scale outflow discovered by Yun \& Clemens (1994) is 
actually composed of two approximately co-oriented pairs of lobes. 
The outflow centers
are $\sim$10\arcsec\ apart and coincide with the 
assumed origins of the two NIR reflection nebulae, 
thus confirming the presence of two embedded protostars (Fig. 3). 
The outflow velocities are small ($\Delta v_{\rm mean} \sim 0.7$\,km/s, 
$\Delta v_{\rm max} \sim 1$\,km/s)
and the exact alignment cannot be derived since the southern, red lobes 
overlap.
The extended $^{13}$CO gas at the rest velocity of the cloud core 
($v_{\rm LSR} = 2.9\pm 0.2$\,km/s) is completely resolved out in the 
interferometer image.

\section{CB\,230 in the framework of binary star formation} 

The globule CB\,230 contains an embedded, wide binary protostellar system 
with a projected separation of $\sim 4500$\,AU, placing the system 
at the upper end of the pre-main sequence binary separation distribution 
(Mathieu 1994). 
The specific angular momentum of the double core is 
$J/M$\,$\sim$\,5$\times$10$^{-17}$\,pc$^2$\,s$^{-1}$, 
which is at the lower end of values for dense cores in dark clouds 
(e.g., Goodman et al. 1993), but still at the upper end of the range 
for pre-main sequence binary star systems (Simon et al. 1995). 
In this sense, the CB\,230 system closes the gap between pre-star-forming 
molecular cloud cores and young binary star systems. 
The ratio of rotational kinetic to gravitational energy of the 
double core is $\beta \sim 0.01$. This value is too small to explain 
the binary formation due to pure rotational fission, but large enough to be 
consistent with 
fragmentation of a slowly rotating, magnetically supported cloud 
(Bodenheimer, Boss, Klein, this volume).

\end{document}